\documentclass[amsmath,superscriptaddress,reprint]{revtex4-1}

\usepackage{graphicx}
\usepackage{hyperref}

\begin{document}
\title
{Single carbon nanotubes as ultrasmall all-optical memories}
\author{T.~Uda}
\affiliation{Nanoscale Quantum Photonics Laboratory, RIKEN, Saitama 351-0198, Japan}
\affiliation{Department of Applied Physics, The University of Tokyo, Tokyo 113-8656, Japan}
\author{A.~Ishii}
\affiliation{Nanoscale Quantum Photonics Laboratory, RIKEN, Saitama 351-0198, Japan}
\affiliation{Quantum Optoelectronics Research Team, RIKEN Center for Advanced Photonics, Saitama 351-0198, Japan}
\author{Y.~K.~Kato}
\email[Corresponding author. ]{yuichiro.kato@riken.jp}
\affiliation{Nanoscale Quantum Photonics Laboratory, RIKEN, Saitama 351-0198, Japan}
\affiliation{Quantum Optoelectronics Research Team, RIKEN Center for Advanced Photonics, Saitama 351-0198, Japan}

\begin{abstract}
Performance improvements are expected from integration of photonic devices into information processing systems, and in particular, all-optical memories provide a key functionality. Scaling down the size of memory elements is desirable for high-density integration, and the use of nanomaterials would allow for devices that are significantly smaller than the operation wavelengths. Here we report on all-optical memory based on individual carbon nanotubes, where adsorbed molecules give rise to optical bistability. By exciting at the high-energy tail of the excitonic absorption resonance, nanotubes can be switched between the desorbed state and the adsorbed state. We demonstrate reversible and reproducible operation of the nanotube optical memory, and determine the rewriting speed by measuring the molecular adsorption and desorption times. Our results underscore the impact of molecular-scale effects on optical properties of nanomaterials, offering new design strategies for photonic devices that are a few orders of magnitude smaller than the optical diffraction limit.
\end{abstract}

\maketitle

On-chip photonic devices can potentially boost the capabilities of modern information-processing systems by replacing their electrical counterparts, as they offer a number of advantages such as low power dissipation, high speed processing, and reduced crosstalk \cite{Caulfield:2010}. High-efficiency photon generation can be achieved with thresholdless lasing in nanoscale metallic cavities \cite{Khajavikhan:2012}, while terahertz modulation has been accomplished using silicon-polymer hybrid waveguides \cite{Hochberg:2006}. Further development towards optical computing requires advances in key devices such as all-optical memories and switches.
These functional devices usually employ optical bistability, where two optically distinguishable states can be selected by optical means \cite{Gibbs:1985}. 
The majority of the optical bistable devices rely on cavity effects \cite{Almeida:2004, Hill:2004, Mori:2006, Liu:2010, Nozaki:2012, Kuramochi:2014}, since they provide strong light confinement for enhancing nonlinearity and allow miniaturization of the systems. 
Even the smallest cavities, however, have length scales of a few microns needed to confine the light waves, putting a limit to size reduction.
Another route for high-density integration of optical memories and switches is the use of nanomaterials that exhibit optical bistability.

Single-walled carbon nanotubes~(CNTs) are appealing for such applications because they can be integrated with silicon photonics \cite{Gaufres:2012, Imamura:2013, Miura:2014} and their optical properties have unique characteristics. With the atomically-thin structures, weak dielectic screening results in tightly bound excitons dominating the optical transitions even at room temperature \cite{Ando:1997, Wang:2005, Maultzsch:2005}. The excitonic energies are structure dependent, offering resonances throughout the telecommunication wavelength \cite{Weisman:2003}, while one-dimensionality gives rise to linearly polarized selection rules \cite{Ajiki:1994}. By using a suitable combination of wavelength and polarization, individual CNTs can be addressed, providing access to ultrasmall volumes \cite{Liu:2013}. Since environmental screening plays an important role in determining the excitonic levels \cite{Ohno:2006, Miyauchi:2007}, suspended CNTs show a considerable modification of the transition energies even with water molecule adsorption \cite{Finnie:2005, Lefebvre:2008, Homma:2013, Xiao:2014}. Excitation-power-dependent photoluminescence (PL) emission energy has been attributed to heating-induced molecular desorption \cite{Milkie:2005, Moritsubo:2010}, suggesting that optical control of CNT emission properties is possible via the adsorbed molecules.

Here we report on optical bistability in individual carbon nanotubes, arising from excitonic resonance shifts induced by molecular adsorption and desorption. Power dependence measurements show that the nanotubes can take two different emission states under the same excitation condition, and the optical bistability is only observed for excitation energies sufficiently higher than the excitonic absorption peak characteristic of CNTs. We find that shifting of the $E_{22}$ resonance can explain the hysteresis, where a large difference in the laser heating efficiency for the two states results in different threshold powers. We employ the bistability to demonstrate reversible and reproducible switching operation, and perform time-resolved measurements to determine the rewriting speed of the nanotube optical memory.

\section*{Results and discussion}

\begin{figure}
\includegraphics{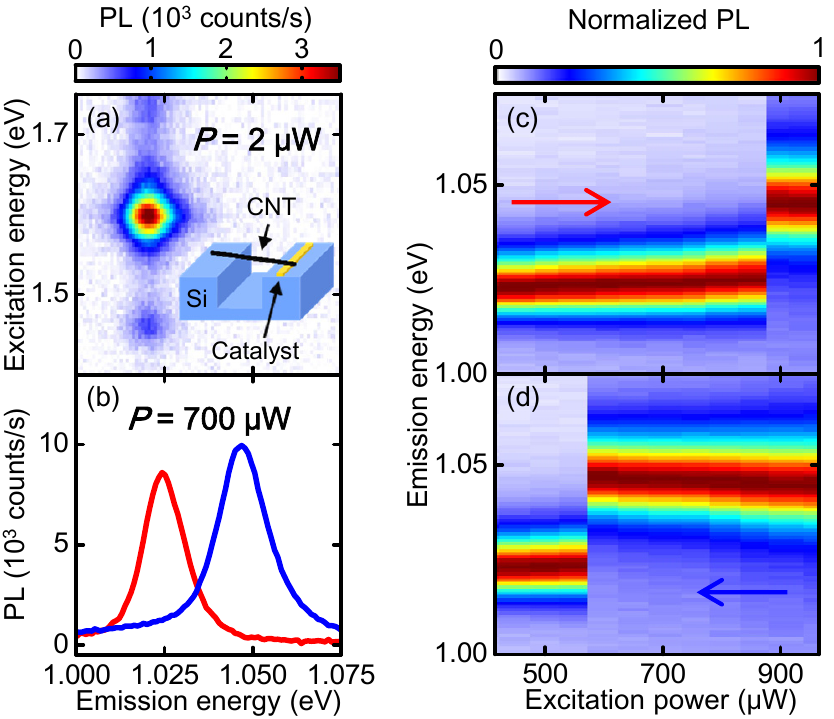}
\caption{\label{fig1}
Optical bistability of an individual CNT.
(a) A PL excitation map of a 2.38-$\mu$m-long (10,5) nanotube taken with $P=2$~$\mu$W, showing an emission peak at 1.021~eV and an excitation resonance at 1.603~eV. The inset is a schematic of a sample.
(b) PL spectra measured at $P=700$~$\mu$W during an up sweep (red curve) and a down sweep (blue curve). 
(c) and (d) Excitation power dependence of normalized PL spectra for an up sweep and a down sweep, respectively. 
The spectra are normalized by the maximum intensity at each power. All data are taken with laser polarization parallel to the tube axis, and $\Delta E_\text{ex}=+41$~meV is used in (b-d).
}\end{figure}

\paragraph*{Optical bistability in carbon nanotubes.}
Our samples are individual CNTs suspended over trenches on silicon substrates (inset of Figure~\ref{fig1}a). The nanotubes are synthesized by alcohol chemical vapor deposition \cite{Maruyama:2002} from catalyst particles placed near the trenches. We then put the samples in a PL microscopy system \cite{Jiang:2015, Ishii:2015}, where they are kept in nitrogen to prevent formation of oxygen-induced defects \cite{Georgi:2008, Yoshikawa:2010}. We note that the samples have been exposed to atmosphere during transfer from the synthesis furnace to the optical system.

A typical PL excitation map of an individual nanotube measured with a laser power $P=2$~$\mu$W is shown in Figure~\ref{fig1}a. Clear resonances of emission and excitation are observed, which correspond to the $E_{11}$ and $E_{22}$ energies, respectively. By comparing these energies to tabulated data \cite{Ishii:2015}, we assign the chirality of the nanotube to be (10, 5).

In Figure~\ref{fig1}b, we show PL spectra of this nanotube taken at $P=700$~$\mu$W during an up sweep (red curve) and a down sweep (blue curve) of the excitation power, where an excitation energy $E_\text{ex}=1.644$~eV is used. The spectrum for the up sweep shows a peak at 1.025~eV, which is close to the emission energy in the PL excitation map. For the down sweep, however, we observe a single peak at a much higher energy of 1.047~eV.

In order to understand the origin of the different emission states, detailed excitation power dependence of PL emission spectra is measured. In the up sweep (Figure~\ref{fig1}c), an abrupt spectral change is observed at $P=870$~$\mu$W, which is attributed to the laser-heating induced molecular desorption \cite{Milkie:2005, Moritsubo:2010}. Interestingly, we find that the abrupt spectral change occurs at a much lower power of $P=570$~$\mu$W for the down sweep (Figure~\ref{fig1}d). The hysteretic behavior immediately indicates the presence of optical bistability \cite{Gibbs:1985}, as the nanotube takes two different emission states depending on its excitation history.

\begin{figure*}
\includegraphics{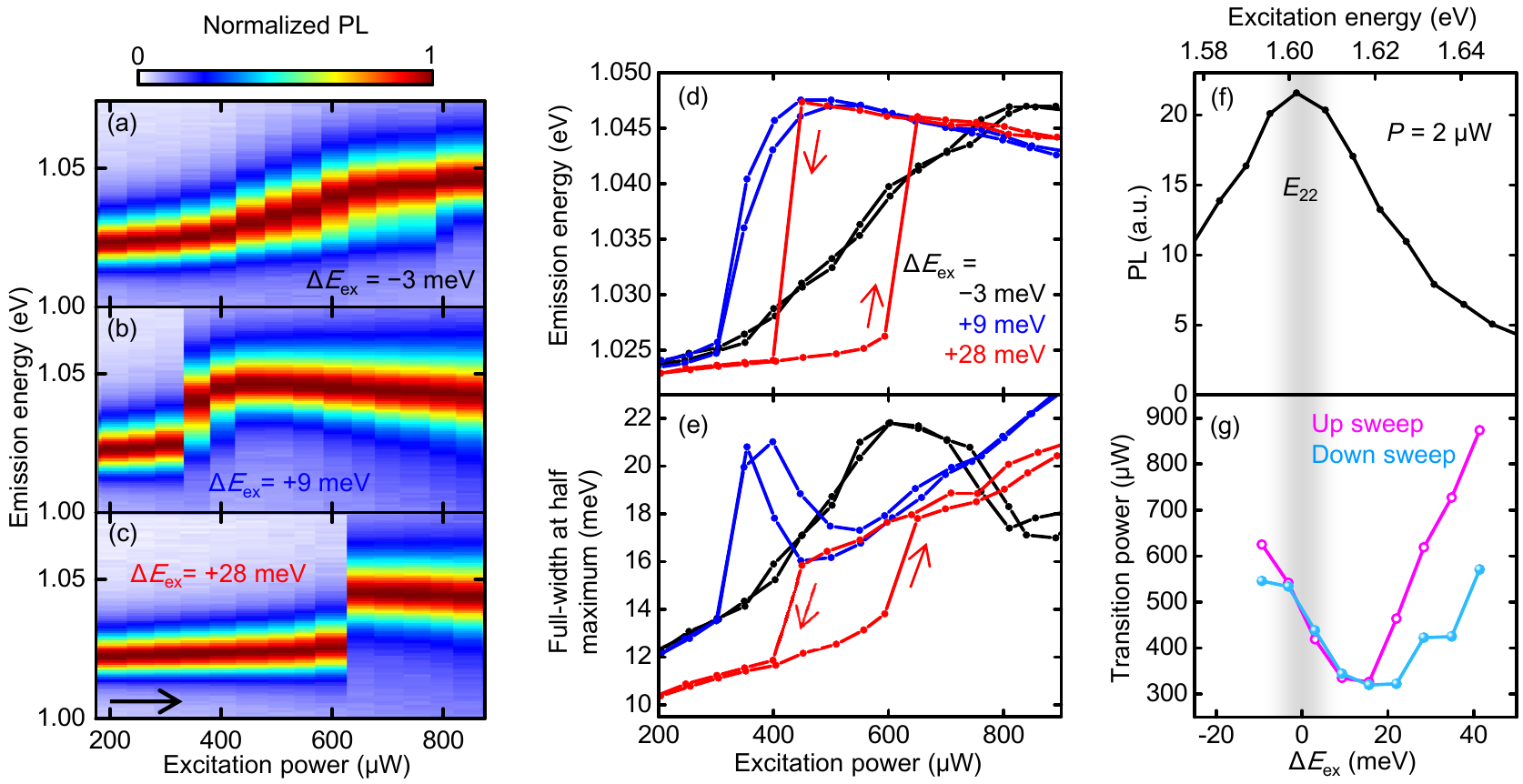}
\caption{\label{fig2}
Detuning energy dependence of the spectral transitions.
Excitation power dependence of the normalized PL spectra taken with (a) $\Delta E_\text{ex}=-3$~meV, 
(b) $\Delta E_\text{ex}=+9$~meV, and (c) $\Delta E_\text{ex}=+28$~meV. The spectra are normalized by the maximum intensity at each power. 
Excitation power dependence of (d) emission energy and (e) full-width at half-maximum, obtained by fitting the PL spectra with a Lorentzian function. The detuning energies of $-3$~meV (black), $+9$~meV (red), and $+28$~meV (blue) are used. Broadening observed around $P=600$~$\mu$W for $\Delta E_\text{ex}=-3$~meV and 
near $P=400$~$\mu$W for +9~meV in (e) are artifacts caused by the overlap of the emission peaks from the adsorbed and desorbed states.
(f) PL excitation spectrum. Lorentzian fits are performed on the emission spectra in Figure~\ref{fig1}a, and the peak area is plotted as a function of excitation energy.
(g) Detuning energy dependence of the transition power for up sweeps (pink open circles) and down sweeps (light-blue filled circles). The transition powers are determined by interpolating the excitation power at an emission energy of 1.035~eV.
The gray shaded bar indicates the $E_{22}$ resonance determined by fitting the PL excitation spectrum in (f) with a Lorentzian function. All data are taken with laser polarization parallel to the tube axis.
}\end{figure*}

We have performed such excitation power dependence measurements for various detuning energies $\Delta E_\text{ex}=E_\text{ex}-E_{22}$ to investigate the mechanism of the bistability, and typical up-sweep data are plotted in Figure~\ref{fig2}a-c. When we excite the nanotube with $\Delta E_\text{ex}=-3$~meV, the spectrum shows a continuous blueshift as the excitation power increases (Figure~\ref{fig2}a). The smooth spectral shift implies that the adsorbed molecules are gradually desorbing by laser-induced heating. As the excitation energy is increased to $\Delta E_\text{ex}=+9$~meV, the spectrum shows a more abrupt blueshift (Figure~\ref{fig2}b), and a discontinuous blueshift is observed at a much higher energy of $\Delta E_\text{ex}=+28$~meV (Figure~\ref{fig2}c).

In Figure~\ref{fig2}d, we show both the up- and down-sweep power dependence of the emission energy measured at these three detuning energies. Hysteresis is not observed at relatively small detunings of $-3$~meV and $+9$~meV, while a clear hysteresis loop opens up for the large positive detuning energy of $+28$~meV. The hysteretic behavior presented in Figure~\ref{fig1}c,d have also been taken with a large detuning of $+41$~meV, suggesting the importance of detuning. 

In fact, the differences in the spectral transitions and the appearance of hysteresis can be understood by taking into account associated shifts of the absorption peak. The molecular desorption causes a blueshift not only for $E_{11}$ but also for the $E_{22}$ resonance \cite{Lefebvre:2008, Homma:2013}. Depending on the detuning, such a blueshift can result in a reduction or an enhancement of the absorption efficiency. 

When we excite the nanotube near the absorption peak with $\Delta E_\text{ex}=-3$~meV, molecular desorption shifts the peak away from the excitation energy. The efficiency of the laser-induced heating is then reduced, which suppresses further desorption of molecules. This negative feedback on molecular desorption stabilizes the amount of adsorbed molecules, and increasing the excitation power results in a smooth blueshift of the PL spectrum. As there is a one-to-one correspondence between the excitation power and the emission energy, hysteresis is absent.

In comparison, exciting at the higher energy tail of the $E_{22}$ resonance with  $\Delta E_\text{ex}=+28$~meV leads to a positive feedback. Once the molecular desorption starts due to laser-induced heating, $E_{22}$ moves closer to the excitation energy by the blueshift. The improved heating efficiency drives additional molecular desorption, which in turn causes a further blueshift of the $E_{22}$ resonance. Because of this positive feedback on molecular desorption, the nanotube transitions from the cold adsorbed state with low heating efficiency to the hot desorbed state with high heating efficiency, which appears as a discontinuous blueshift at the threshold power.
Considerable broadening of the desorbed state is consistent with this picture (Figure~\ref{fig2}e), since the tube temperature is known to be proportional to the width of the emission peak \cite{Lefebvre:2004, Matsuda:2008_2, Yoshikawa:2009}.

The large difference in the heating efficiency for the cold adsorbed state and the hot desorbed state gives rise to the optical hysteresis. For the up sweep taken with $\Delta E_\text{ex}=+28$~meV~(Figure~\ref{fig2}d), $P\sim600$~$\mu$W is required for the cold desorbed state to reach the transition temperature because of the low heating efficiency. When we sweep the excitation power down, the nanotube stays in the hot desorbed state which has a higher heating efficiency, resulting in a transition at a much lower power of $\sim$400~$\mu$W compared to the up sweep.

We further investigate optical bistability at larger detuning energies. The PL excitation spectrum (Figure~\ref{fig2}f) shows that the high energy tail of the $E_{22}$ peak is steep, and thus the transition will occur at a significantly larger power when the detuning energy is increased by just tens of meV. In Figure~\ref{fig2}g, the transition powers for the adsorbed and desorbed states obtained from the up and down sweeps, respectively, are plotted as a function of the detuning energy. The transition power of the adsorbed state shows a rapid increase as we expect, while the desorbed state shows a gradual increase. Such a difference can be explained by the broader absorption peak of the hot desorbed state compared to the cold adsorbed state, as the laser absorption efficiency would be less sensitive to the excitation energy. The discrepancy in the detuning energy dependence results in an increase of the width of the hysteresis, or equivalently the bistable power region, for larger $\Delta E_\text{ex}$.

\begin{figure}
\includegraphics{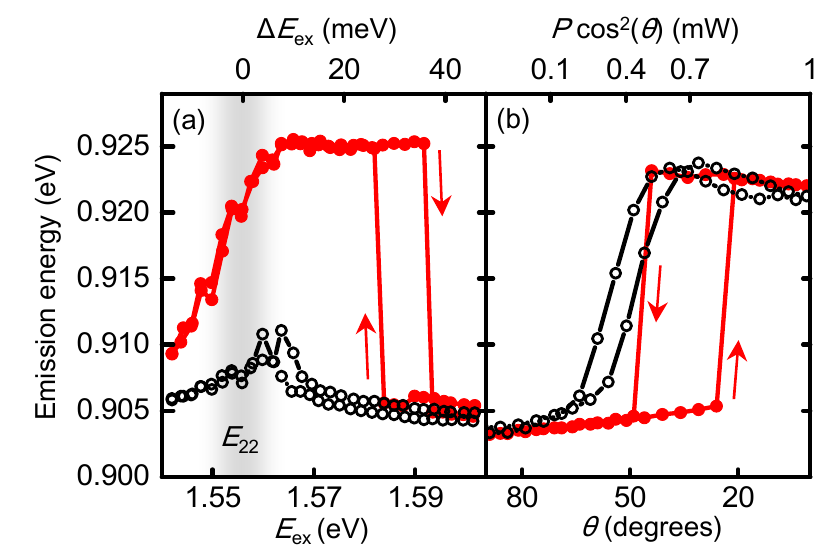}
\caption{\label{fig3}
Bistability observed in excitation energy and polarization dependence for a (9,8) tube with a length of 2.01~$\mu$m.
(a) Excitation energy dependence of the emission energy measured with $P=500$~$\mu$W (red filled circles) and 250~$\mu$W (black open circles), taken with polarization parallel to the tube axis. The gray shaded bar indicates the $E_{22}$ resonance for this nanotube determined from a PL excitation spectrum taken with the same polarization at $P=5$~$\mu$W.
(b) Polarization dependence of the emission energy measured with $\Delta E_\text{ex}=+34$~meV (red filled circles) and +4~meV (black open circles), taken at $P=1.00$~mW. For both panels, the emission energies are determined by fitting the spectra with a Lorentzian function.
}\end{figure}

The results shown in Figure~\ref{fig2}g suggest that the bistability can also be observed when the PL emission energies are measured as a function of detuning energy. Red filled circles in Figure~\ref{fig3}a shows the excitation energy dependence taken at  $P=500$~$\mu$W for a (9,8) nanotube. When the excitation energy is swept in a direction approaching $E_{22}$ from higher energy, a discontinuous blueshift of the emission energy occurs at $\Delta E_\text{ex}\sim+25$~meV. When the excitation energy is swept in the opposite direction, as expected, we observe an abrupt redshift at a higher transition energy of $\Delta E_\text{ex}\sim+35$~meV. It is also noteworthy that the emission energy shows a gradual shift when $\Delta E_\text{ex}\lesssim0$~meV, because the negative feedback for molecular desorption occurs at sufficiently small or negative detuning energies. When the power is lowered to 250~$\mu$W (black open circles in Figure~\ref{fig3}a), the emission energy transitions are not observed anymore as the tube temperature does not reach the threshold.

In addition to the excitation energy and power dependence, the one-dimensional structure of CNTs gives rise to bistability in polarization angle dependence measurements. The absorption of an excitation beam perpendicularly polarized to the tube axis is strongly suppressed due to the depolarization effect \cite{Ajiki:1994}, and therefore the beam polarized along the tube axis is predominantly absorbed. When we let the laser polarization angle from the tube axis to be $\theta$, the absorbed power can be approximated by $P\cos^{2} \theta$. The polarization dependence should then be similar to the power dependence in Figure~\ref{fig2}d. Indeed, a large optical hysteresis is observed for an excitation with $\Delta E_\text{ex}=+34$~meV (red filled circles in Figure~\ref{fig3}b), but not for a smaller detuning $\Delta E_\text{ex}=+4$~meV (black open circles in Figure~\ref{fig3}b). 

\paragraph*{Reversible and reproducible optical memory operation.}
The optical hysteresis loops can be exploited to achieve memory operation, since the two stable emission states reflect their excitation history. During an up-sweep power dependence measurement, the nanotube stays in the adsorbed state as long as the power is within the bistable region. Switching to the desorbed state occurs at the transition power, and the nanotube remains in this state during a down sweep until the power is reduced below the bistable region. It should be possible to sustain the emission state by exciting at the power within the bistable region, whereas appropriate modulation of the excitation power would allow for switching between the states.

\begin{figure}
\includegraphics{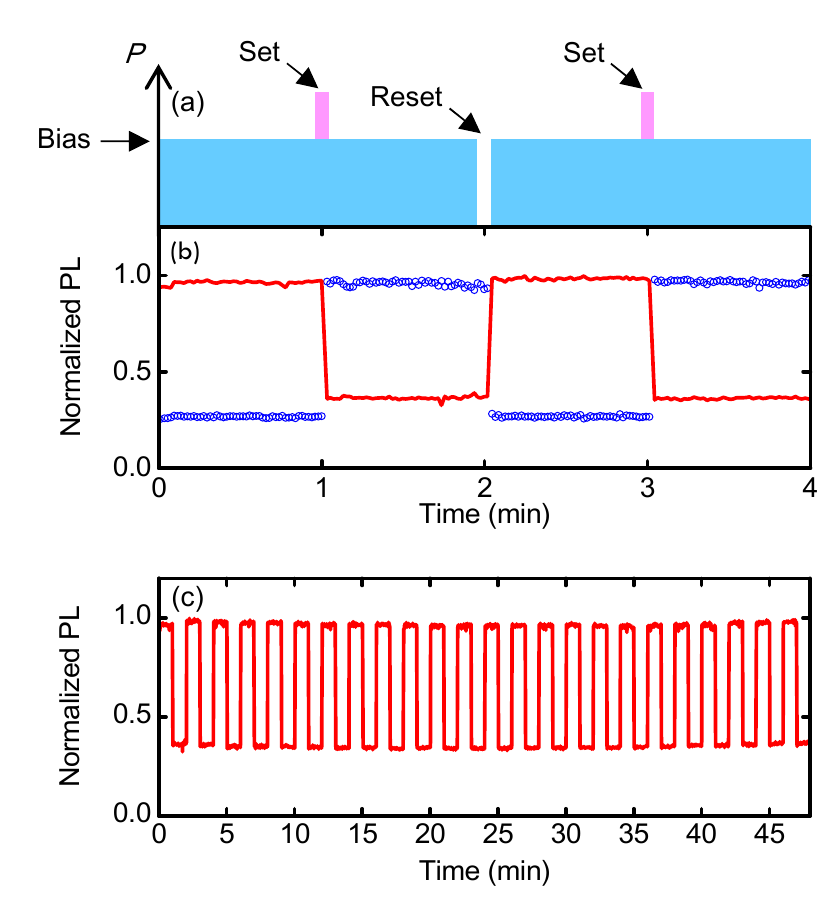}
\caption{\label{fig4}
Reversible and reproducible memory operation of a 2.00-$\mu$m-long (9,8) tube.
(a) Schematic showing the temporal pattern of excitation power. 
(b) Temporal evolution of integrated PL intensities for emission energies of 0.904~eV (red line) and 0.921~eV (blue open circles). The spectral integration windows are 4~meV wide, and the intensities are normalized by the maximum values. In this measurement, the detuning energy is $+42$~meV. The bias beam is tuned to be $P=1.70$~mW and $\theta=-45$$^\circ$, and the set pulses are set to $P=0.89$~mW and $\theta=+45$$^\circ$.
(c) Further evolution of integrated PL at 0.904~eV obtained by continuing the measurement in (b).
}\end{figure}

To verify whether CNTs can actually operate as a rewritable optical memory, we use an excitation pattern shown in Figure~\ref{fig4}a and measure the temporal changes of PL for a different (9,8) nanotube. As the adsorbed and desorbed states are characterized by lower and higher emission energies, respectively, we plot the corresponding temporal evolutions of the spectrally integrated PL for these two energies (Figure~\ref{fig4}b). 

In this measurement, the power of the bias beam is tuned to the center of the bistable power region of the nanotube, while the power of the set pulse is adjusted so that it can induce a transition. At the start of the measurement, a shutter is opened to begin exciting the nanotube with the bias beam. The low-energy PL intensity (red line in Figure~\ref{fig4}b) is large while the high-energy PL intensity (blue open circles in Figure~\ref{fig4}b) is small, indicating that the nanotube is initially in the adsorbed state. One minute after the start of the measurement, a set pulse is added to the bias beam by opening another shutter for a less than a second. The set pulse increases the temperature of the nanotube and switches its state from the adsorbed state to the desorbed state. The switching behavior can be seen as an abrupt increase of the high-energy PL intensity as well as a decrease of the low-energy PL intensity. After another minute from the set operation, the bias beam is blocked by the shutter to perform the reset operation. The nanotube switches back to the adsorbed state and the low-energy PL intensity recovers. We note that the nanotube maintains its emission state during the one-minute intervals between the set and reset operations, showing that the width of the bistable power region is sufficiently large compared to the fluctuations of the absorbed laser power. The results of this measurement confirm the rewritability and the stability of the single CNT optical memory.

We further repeated the switching cycles to check the reliability of the memory operation. In Figure~\ref{fig4}c, we plot the temporal evolution for the low-energy PL intensity. Reversible and reproducible switching operations of over 45 times are demonstrated, suggesting that the nanotube optical memory can also be manipulated under a more complicated sequence of light pulses.

\paragraph*{Desorption times.} 

\begin{figure}
\includegraphics{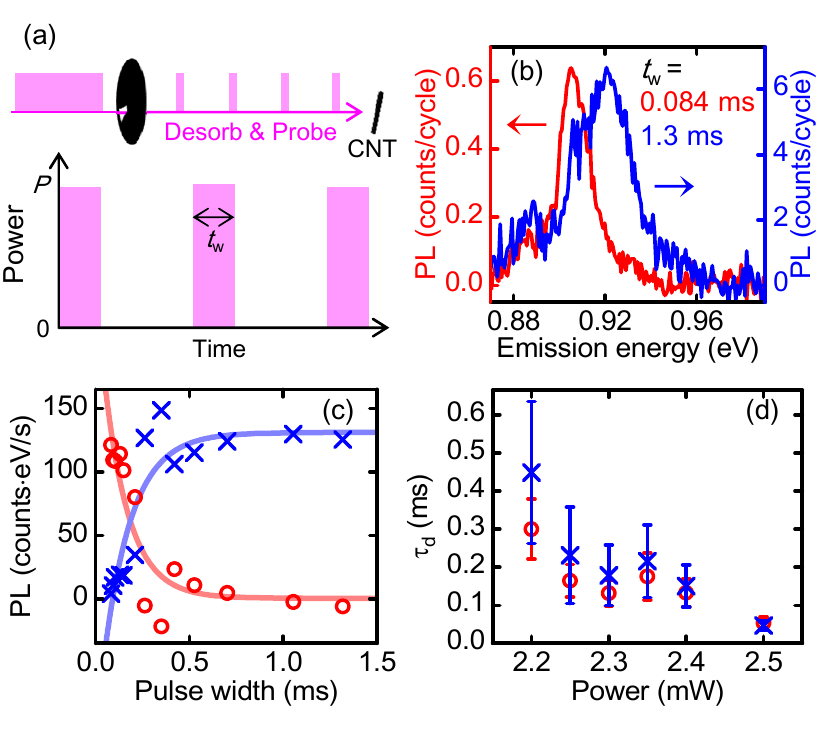}
\caption{\label{fig5}
Desorption time measurements.
(a) Schematic of the measurements showing an excitation pulse pattern and the definition of $t_\text{w}$.
(b) Time-integrated PL spectra taken with $t_\text{w}=0.084$~ms (red curve, left axis) and 1.3~ms (blue curve, right axis).
(c) Transient PL intensities of the high-energy peak (blue crosses) and the low-energy peak (red open circles). Blue and red curves are fits as explained in the text.
(d) Pulse power dependence of $\tau_\text{d}$. 
The transient PL intensities of the high-energy peak and the low-energy peak are used to obtain the blue crosses and the red open circles, respectively. (b-d) are taken with $\Delta E_\text{ex}=+40$~meV and $\theta=+45$$^\circ$ on a 2.62-$\mu$m-long (9,8) tube, and (b,c) are measured at $P=2.40$~mW.
}\end{figure}

Rewriting speed is an important factor in the performance of optical memories. In the case of our nanotube optical memory, the time scale should be limited by the molecular desorption or adsorption times. We start with the investigation of the molecular desorption times by measuring the changes of the PL spectra as a function of the excitation pulse width $t_\text{w}$ (Figure~\ref{fig5}a), using a power high enough to desorb the molecules.

The measurements have been performed on another (9,8) tube, and two representative time-integrated PL spectra are shown in Figure~\ref{fig5}b. The spectrum taken with $t_\text{w}=1.3$~ms shows a large contribution from the blueshifted desorbed-state peak at 0.921~eV, which is reasonable because all molecules should have desorbed for sufficiently long  $t_\text{w}$. In the case of the spectrum taken with $t_\text{w}=0.084$~ms, however, the adsorbed-state peak at 0.906~eV is pronounced. The single peak at a lower-energy shows that most of the molecules stay adsorbed during the pulse, indicating that $t_\text{w}$ is much shorter than the desorption time.

In order to extract the desorption time $\tau_\text{d}$, we analyze the time-integrated PL spectra taken for various pulse widths. Bi-Lorentzian fits to the spectra are performed to obtain the desorbed-state and adsorbed-state peak areas. We then numerically differentiate the peak area by $t_\text{w}$ to examine the transient behavior of the nanotube during the excitation pulse (Figure~\ref{fig5}c). The intensity of PL from the desorbed state (blue crosses) increases as the pulse width becomes wider, while the intensity of PL from the adsorbed state (red open circles) decreases. Both intensities reach constant values at longer pulse widths, showing that all molecules have desorbed. We fit the transient PL intensities of the desorbed state and the adsorbed state with $I_\text{PL}= I_\text{0}[1-\exp(-(t_\text{w}-t_\text{0})/\tau_\text{d})]$ and $I_\text{PL}=I_\text{1}\exp(-(t_\text{w}-t_\text{0})/\tau_\text{d})$, respectively, where $I_\text{0}$ is the saturation intensity, $I_\text{1}$ is the initial intensity, and $t_\text{0}$ is the offset time. Solid curves in Figure~\ref{fig5}c are the best fits to the data, where we obtain comparable values of $\tau_\text{d}=0.15$~ms (blue curve) and 0.13~ms (red curve). 

The desorption times are known to become shorter for higher material temperatures \cite{Adamson:1990}, and therefore we expect $\tau_\text{d}$ to decrease for larger excitation powers. We have taken the $t_\text{w}$ dependence of the spectra at various powers and the results are summarized in Figure~\ref{fig5}d. Indeed, $\tau_\text{d}$ becomes shorter as the power increases, indicating that the molecules do desorb faster.

\begin{figure}
\includegraphics{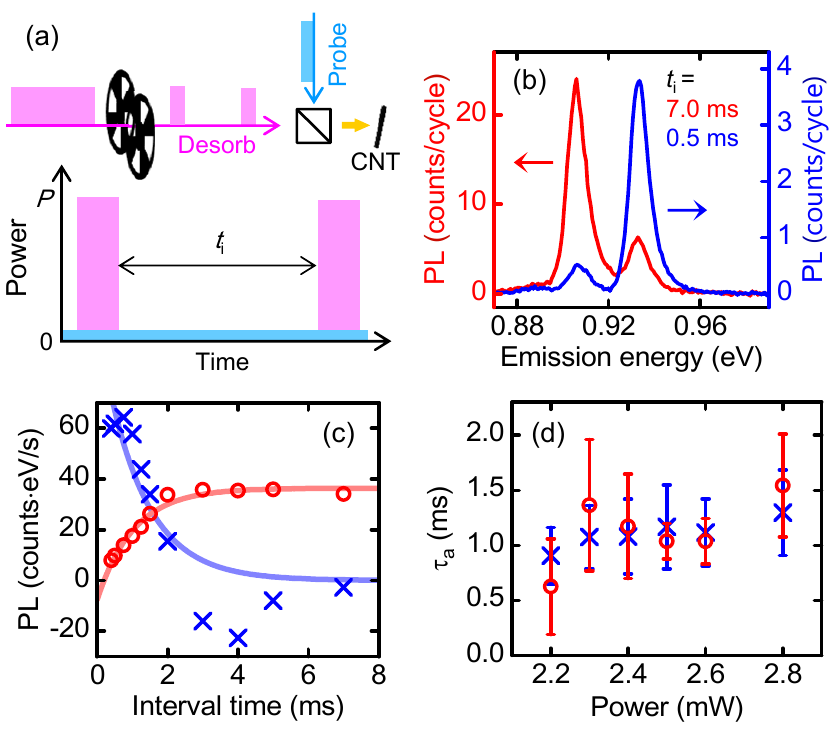}
\caption{\label{fig6}
Adsorption time measurements.
(a) Schematic of the measurements showing desorption pulses and a probe beam as well as the definition of  $t_\text{i}$. 
(b) PL difference spectra taken with $t_\text{i}=7.0$~ms (red curve, left axis) and 0.5~ms (blue curve, right axis).
(c) Transient PL intensities of the low-energy peak (red open circles) and the high-energy peak (blue crosses). Red and blue curves are fits as explained in the text.
(b,c) are measured at $P=2.50$~mW.
(d) Pulse power dependence of $\tau_\text{a}$. 
The transient PL intensities of the low-energy peak and the high-energy peak are used to obtain the red open circles and the blue crosses, respectively. 
All data are taken with $\Delta E_\text{ex}=+40$~meV, and the power ratio of the desorption pulses to the probe beam is adjusted to be 1:0.017. The desorption pulses and probe beam are set to $\theta=+45$$^\circ$ and $-45$$^\circ$, respectively. 
}\end{figure}

\paragraph*{Adsorption times.}
The measurements for molecular adsorption times are not as straightforward as the desorption-time measurements, because adsorption occurs in between the pulses when nanotubes are not emitting. We therefore introduce a weak continuous beam for probing the amount of adsorbed molecules during the pulse interval time $t_\text{i}$ (Figure~\ref{fig6}a). The probe beam power is kept sufficiently low to minimize laser-induced heating, while the desorption pulses are set to high enough powers. As the desorption pulses inevitably cause strong PL, we take the difference between the time-integrated spectra with and without the probe beam to examine the adsorption dynamics during the pulse intervals. 

Figure~~\ref{fig6}b shows two PL difference spectra taken at long ($t_\text{i}=7.0$~ms, red curve) and short ($t_\text{i}=0.5$~ms, blue curve) intervals. In these spectra, two distinct peaks at 0.906~eV and 0.933~eV are observed, which correspond to the adsorbed and desorbed states, respectively. In the spectrum taken at $t_\text{i}=7.0$~ms, the low-energy peak dominates the spectrum, showing that the molecules have already adsorbed. In comparison, the peak at the higher energy is prominent in the spectrum measured at $t_\text{i}=0.5$~ms, indicating that molecular adsorption is insubstantial at this short interval time.

We note that the higher energy peaks in Figure~\ref{fig6}b and Figure~\ref{fig5}b are located at slightly different energies, although both peaks should correspond to the desorbed state. 
Such a difference is reasonable, as the nanotube is probed with significantly higher powers in the desorption-time measurements compared to the adsorption-time measurements. 
Since the temperature of the nanotube should be higher, heating-induced redshift is expected \cite{Lefebvre:2004}. In comparison, the lower-energy peaks are at the same energy in Figure~\ref{fig6}b and Figure~\ref{fig5}b, indicating that the heating effect is negligible for the adsorbed state.

We further examine the adsorption dynamics by analyzing the inter-pulse emission using $t_\text{i}$ dependence of the PL difference spectra. As in the case for the desorption-time measurements, the emission spectra are fitted with a bi-Lorentzian function to obtain the peak areas, and we numerically differentiate the peak areas by $t_\text{i}$ to evaluate the transient emission states during the pulse interval (Figure~\ref{fig6}c). The PL intensity from the adsorbed state increases for longer $t_\text{i}$ (red open circles), while emission from the desorbed state decreases (blue crosses).

To extract the molecular adsorption time $\tau_\text{a}$, $I_\text{PL}= I_\text{1}[1-\exp(-(t_\text{i}-t_\text{0})/\tau_\text{a})]$ and $I_\text{PL}=I_\text{0}\exp(-(t_\text{i}-t_\text{0})/\tau_\text{a})$ are fit to the adsorbed state and the desorbed state emission intensities, respectively. The best fitted curves are shown in Figure~\ref{fig6}c, and similar values of $\tau_\text{a}=1.0$~ms (red curve) and 1.2~ms (blue curve) are obtained. The adsorption time is about an order of magnitude longer than the desorption times, limiting the rewriting speed of the nanotube optical memories.

We have also investigated the desorption-pulse power dependence of adsorption time, and the results are summarized in Figure~\ref{fig6}d. We do not observe a clear power dependence for $\tau_\text{a}$, although measurements up to higher powers have been performed compared to the $\tau_\text{d}$ measurements. The absence of power dependence is expected, since the adsorption time only depends on the molecular collision frequency if the sample cooling time does not limit the experimental resolution \cite{Berber:2000, Fujii:2005, Yu:2005}. It should be possible to shorten the adsorption time by increasing the humidity, as the collision frequency is determined by the partial pressure of the molecules. 

\section*{Conclusions}
We observe optical bistability in individual CNTs, where switching between the hot desorbed state and the cold adsorbed state is triggered by a change in the excitation laser power. A direct readout of the nanotube state is possible, as the emission energy switches simultaneously. The $E_{22}$ resonance shift associated with molecular adsorption causes a considerable difference in the laser heating efficiency, locking the nanotube into either of the two states. Furthermore, we demonstrate reversible and reproducible optical memory operation, and obtain the adsorption and desorption times by performing time-resolved measurements.

Our results highlight the potential use of CNTs for ultrasmall optical memories and switches in photonic circuits, surpassing the size limits imposed by the cavities.
Taking advantage of various chiralities and polarization selectivity, it should be possible to address individual nanotubes even if they are located in the vicinity.
The functionality of the nanotube optical memory is provided by the few molecules on the surface, which is coupled to the characteristic absorption peak through the strong effects of screening.
Being telecom-band emitters that can be integrated with silicon photonics, CNTs present opportunities for optical devices with operation mechanisms at the molecular level.

\section*{Methods}

\paragraph*{Photoluminescence microscopy.}
The PL measurements are carried out with a home-built confocal microspectroscopy system \cite{Jiang:2015, Ishii:2015}. A continuous-wave Ti:sapphire laser is used for excitation, and an objective lens with a numerical aperture of 0.8 and a working distance of 3.4~mm focuses the laser onto the sample.  The linear polarization of the excitation laser is rotated using a half-wave plate and PL from the sample is detected by an InGaAs photodiode array detector attached to a spectrometer. All measurements are conducted at room temperature in nitrogen.

\paragraph*{Carbon nanotubes suspended over trenches.} 
We synthesize CNTs on Si chips with prefabricated trenches, and characterize the nanotubes by PL microscopy \cite{Ishii:2015}. Electron beam lithography and dry etching processes are performed on Si substrates to form the trenches, and catalyst areas are patterned by an additional electron beam lithography step. Fe(III) acetylacetonate and fumed silica dispersed in ethanol are spin-coated as catalysts, and single-walled carbon nanotubes are grown over the trenches by alcohol chemical vapor deposition. We identify the nanotube chirality by PL excitation spectroscopy, and select bright nanotubes for our experiments.

\paragraph*{Demonstration of optical memory operation.}
A polarizing beam splitter is used for separating the laser beam into two orthogonal polarizations, and the ratio of the two beams is controlled by rotating the polarization before the beam splitter. The transmitted beam is used for the bias beam, and the reflected beam is used for the set pulses by placing a shutter in its path. We use another shutter before the beam splitter to perform reset operations. The two beams are directed towards the objective lens using a second polarizing beam splitter, and we rotate the polarization of the beams to be $\theta=\pm45$$^\circ$ with a half-wave plate placed after this beam splitter. We note that the power absorbed by the nanotubes is therefore half of the power $P$ measured at the sample.

\paragraph*{Time-resolved measurements.}
Optical choppers are used to obtain optical pulses. In desorption-time measurements, we have stacked two chopper wheels to achieve a duty cycle of 0.0053. Even for the highest frequency used in our measurements, the pulse interval is longer than 16~ms due to the low duty cycle, which is still long enough for re-adsorption of the molecules. In the adsorption-time measurements, two optical choppers are synchronized to the same reference. The frequency and the phase difference are controlled to vary $t_\text{i}$ while keeping $t_\text{w}=0.4$~ms.

\begin{acknowledgments}
Work supported by JSPS (KAKENHI JP16H05962) and MEXT (Photon Frontier Network Program, Nanotechnology Platform). T.U. is supported by ALPS and JSPS Research Fellowship.
\end{acknowledgments}


\begin{thebibliography}{39}%
\makeatletter
\providecommand \@ifxundefined [1]{%
 \@ifx{#1\undefined}
}%
\providecommand \@ifnum [1]{%
 \ifnum #1\expandafter \@firstoftwo
 \else \expandafter \@secondoftwo
 \fi
}%
\providecommand \@ifx [1]{%
 \ifx #1\expandafter \@firstoftwo
 \else \expandafter \@secondoftwo
 \fi
}%
\providecommand \natexlab [1]{#1}%
\providecommand \enquote  [1]{``#1''}%
\providecommand \bibnamefont  [1]{#1}%
\providecommand \bibfnamefont [1]{#1}%
\providecommand \citenamefont [1]{#1}%
\providecommand \href@noop [0]{\@secondoftwo}%
\providecommand \href [0]{\begingroup \@sanitize@url \@href}%
\providecommand \@href[1]{\@@startlink{#1}\@@href}%
\providecommand \@@href[1]{\endgroup#1\@@endlink}%
\providecommand \@sanitize@url [0]{\catcode `\\12\catcode `\$12\catcode
  `\&12\catcode `\#12\catcode `\^12\catcode `\_12\catcode `\%12\relax}%
\providecommand \@@startlink[1]{}%
\providecommand \@@endlink[0]{}%
\providecommand \url  [0]{\begingroup\@sanitize@url \@url }%
\providecommand \@url [1]{\endgroup\@href {#1}{\urlprefix }}%
\providecommand \urlprefix  [0]{URL }%
\providecommand \Eprint [0]{\href }%
\providecommand \doibase [0]{http://dx.doi.org/}%
\providecommand \selectlanguage [0]{\@gobble}%
\providecommand \bibinfo  [0]{\@secondoftwo}%
\providecommand \bibfield  [0]{\@secondoftwo}%
\providecommand \translation [1]{[#1]}%
\providecommand \BibitemOpen [0]{}%
\providecommand \bibitemStop [0]{}%
\providecommand \bibitemNoStop [0]{.\EOS\space}%
\providecommand \EOS [0]{\spacefactor3000\relax}%
\providecommand \BibitemShut  [1]{\csname bibitem#1\endcsname}%
\let\auto@bib@innerbib\@empty
\bibitem [{\citenamefont {Caulfield}\ and\ \citenamefont
  {Dolev}(2010)}]{Caulfield:2010}%
  \BibitemOpen
  \bibfield  {author} {\bibinfo {author} {\bibfnamefont {H.~J.}\ \bibnamefont
  {Caulfield}}\ and\ \bibinfo {author} {\bibfnamefont {S.}~\bibnamefont
  {Dolev}},\ }\bibfield  {title} {\enquote {\bibinfo {title} {Why future
  supercomputing requires optics},}\ }\href
  {http://dx.doi.org/10.1038/nphoton.2010.94} {\bibfield  {journal} {\bibinfo
  {journal} {Nat. Photonics}\ }\textbf {\bibinfo {volume} {4}},\ \bibinfo
  {pages} {261--263} (\bibinfo {year} {2010})}\BibitemShut {NoStop}%
\bibitem [{\citenamefont {Khajavikhan}\ \emph {et~al.}(2012)\citenamefont
  {Khajavikhan}, \citenamefont {Simic}, \citenamefont {Katz}, \citenamefont
  {Lee}, \citenamefont {Slutsky}, \citenamefont {Mizrahi}, \citenamefont
  {Lomakin},\ and\ \citenamefont {Fainman}}]{Khajavikhan:2012}%
  \BibitemOpen
  \bibfield  {author} {\bibinfo {author} {\bibfnamefont {M.}~\bibnamefont
  {Khajavikhan}}, \bibinfo {author} {\bibfnamefont {A.}~\bibnamefont {Simic}},
  \bibinfo {author} {\bibfnamefont {M.}~\bibnamefont {Katz}}, \bibinfo {author}
  {\bibfnamefont {J.~H.}\ \bibnamefont {Lee}}, \bibinfo {author} {\bibfnamefont
  {B.}~\bibnamefont {Slutsky}}, \bibinfo {author} {\bibfnamefont
  {A.}~\bibnamefont {Mizrahi}}, \bibinfo {author} {\bibfnamefont
  {V.}~\bibnamefont {Lomakin}}, \ and\ \bibinfo {author} {\bibfnamefont
  {Y.}~\bibnamefont {Fainman}},\ }\bibfield  {title} {\enquote {\bibinfo
  {title} {Thresholdless nanoscale coaxial lasers},}\ }\href
  {http://dx.doi.org/10.1038/nature10840} {\bibfield  {journal} {\bibinfo
  {journal} {Nature}\ }\textbf {\bibinfo {volume} {482}},\ \bibinfo {pages}
  {204--207} (\bibinfo {year} {2012})}\BibitemShut {NoStop}%
\bibitem [{\citenamefont {Hochberg}\ \emph {et~al.}(2006)\citenamefont
  {Hochberg}, \citenamefont {Baehr-Jones}, \citenamefont {Wang}, \citenamefont
  {Shearn}, \citenamefont {Harvard}, \citenamefont {Luo}, \citenamefont {Chen},
  \citenamefont {Shi}, \citenamefont {Lawson}, \citenamefont {Sullivan},
  \citenamefont {Jen}, \citenamefont {Dalton},\ and\ \citenamefont
  {Scherer}}]{Hochberg:2006}%
  \BibitemOpen
  \bibfield  {author} {\bibinfo {author} {\bibfnamefont {M.}~\bibnamefont
  {Hochberg}}, \bibinfo {author} {\bibfnamefont {T.}~\bibnamefont
  {Baehr-Jones}}, \bibinfo {author} {\bibfnamefont {G.}~\bibnamefont {Wang}},
  \bibinfo {author} {\bibfnamefont {M.}~\bibnamefont {Shearn}}, \bibinfo
  {author} {\bibfnamefont {K.}~\bibnamefont {Harvard}}, \bibinfo {author}
  {\bibfnamefont {J.}~\bibnamefont {Luo}}, \bibinfo {author} {\bibfnamefont
  {B.}~\bibnamefont {Chen}}, \bibinfo {author} {\bibfnamefont {Z.}~\bibnamefont
  {Shi}}, \bibinfo {author} {\bibfnamefont {R.}~\bibnamefont {Lawson}},
  \bibinfo {author} {\bibfnamefont {P.}~\bibnamefont {Sullivan}}, \bibinfo
  {author} {\bibfnamefont {A.~K.~Y.}\ \bibnamefont {Jen}}, \bibinfo {author}
  {\bibfnamefont {L.}~\bibnamefont {Dalton}}, \ and\ \bibinfo {author}
  {\bibfnamefont {A.}~\bibnamefont {Scherer}},\ }\bibfield  {title} {\enquote
  {\bibinfo {title} {Terahertz all-optical modulation in a silicon-polymer
  hybrid system},}\ }\href {http://dx.doi.org/10.1038/nmat1719} {\bibfield
  {journal} {\bibinfo  {journal} {Nat. Mater.}\ }\textbf {\bibinfo {volume}
  {5}},\ \bibinfo {pages} {703--709} (\bibinfo {year} {2006})}\BibitemShut
  {NoStop}%
\bibitem [{\citenamefont {Gibbs}(1985)}]{Gibbs:1985}%
  \BibitemOpen
  \bibfield  {author} {\bibinfo {author} {\bibfnamefont {H.}~\bibnamefont
  {Gibbs}},\ }\href@noop {} {\emph {\bibinfo {title} {Optical Bistability:
  Controlling Light with Light}}}\ (\bibinfo  {publisher} {Academic Press},\
  \bibinfo {address} {Orlando},\ \bibinfo {year} {1985})\BibitemShut {NoStop}%
\bibitem [{\citenamefont {Almeida}\ and\ \citenamefont
  {Lipson}(2004)}]{Almeida:2004}%
  \BibitemOpen
  \bibfield  {author} {\bibinfo {author} {\bibfnamefont {V.~R.}\ \bibnamefont
  {Almeida}}\ and\ \bibinfo {author} {\bibfnamefont {M.}~\bibnamefont
  {Lipson}},\ }\bibfield  {title} {\enquote {\bibinfo {title} {Optical
  bistability on a silicon chip},}\ }\href
  {http://ol.osa.org/abstract.cfm?URI=ol-29-20-2387} {\bibfield  {journal}
  {\bibinfo  {journal} {Opt. Lett.}\ }\textbf {\bibinfo {volume} {29}},\
  \bibinfo {pages} {2387--2389} (\bibinfo {year} {2004})}\BibitemShut {NoStop}%
\bibitem [{\citenamefont {Hill}\ \emph {et~al.}(2004)\citenamefont {Hill},
  \citenamefont {Dorren}, \citenamefont {de~Vries}, \citenamefont {Leijtens},
  \citenamefont {den Besten}, \citenamefont {Smalbrugge}, \citenamefont {Oei},
  \citenamefont {Binsma}, \citenamefont {Khoe},\ and\ \citenamefont
  {Smit}}]{Hill:2004}%
  \BibitemOpen
  \bibfield  {author} {\bibinfo {author} {\bibfnamefont {M.~T.}\ \bibnamefont
  {Hill}}, \bibinfo {author} {\bibfnamefont {H.~J.~S.}\ \bibnamefont {Dorren}},
  \bibinfo {author} {\bibfnamefont {T.}~\bibnamefont {de~Vries}}, \bibinfo
  {author} {\bibfnamefont {X.~J.~M.}\ \bibnamefont {Leijtens}}, \bibinfo
  {author} {\bibfnamefont {J.~H.}\ \bibnamefont {den Besten}}, \bibinfo
  {author} {\bibfnamefont {B.}~\bibnamefont {Smalbrugge}}, \bibinfo {author}
  {\bibfnamefont {Y.-S.}\ \bibnamefont {Oei}}, \bibinfo {author} {\bibfnamefont
  {H.}~\bibnamefont {Binsma}}, \bibinfo {author} {\bibfnamefont {G.-D.}\
  \bibnamefont {Khoe}}, \ and\ \bibinfo {author} {\bibfnamefont {M.~K.}\
  \bibnamefont {Smit}},\ }\bibfield  {title} {\enquote {\bibinfo {title} {A
  fast low-power optical memory based on coupled micro-ring lasers},}\ }\href
  {http://dx.doi.org/10.1038/nature03045} {\bibfield  {journal} {\bibinfo
  {journal} {Nature}\ }\textbf {\bibinfo {volume} {432}},\ \bibinfo {pages}
  {206--209} (\bibinfo {year} {2004})}\BibitemShut {NoStop}%
\bibitem [{\citenamefont {Mori}, \citenamefont {Yamayoshi},\ and\ \citenamefont
  {Kawaguchi}(2006)}]{Mori:2006}%
  \BibitemOpen
  \bibfield  {author} {\bibinfo {author} {\bibfnamefont {T.}~\bibnamefont
  {Mori}}, \bibinfo {author} {\bibfnamefont {Y.}~\bibnamefont {Yamayoshi}}, \
  and\ \bibinfo {author} {\bibfnamefont {H.}~\bibnamefont {Kawaguchi}},\
  }\bibfield  {title} {\enquote {\bibinfo {title} {Low-switching-energy and
  high-repetition-frequency all-optical flip-flop operations of a polarization
  bistable vertical-cavity surface-emitting laser},}\ }\href {\doibase
  10.1063/1.2181192} {\bibfield  {journal} {\bibinfo  {journal} {Appl. Phys.
  Lett.}\ }\textbf {\bibinfo {volume} {88}},\ \bibinfo {pages} {101102}
  (\bibinfo {year} {2006})}\BibitemShut {NoStop}%
\bibitem [{\citenamefont {Liu}\ \emph {et~al.}(2010)\citenamefont {Liu},
  \citenamefont {Kumar}, \citenamefont {Huybrechts}, \citenamefont {Spuesens},
  \citenamefont {Roelkens}, \citenamefont {Geluk}, \citenamefont {de~Vries},
  \citenamefont {Regreny}, \citenamefont {Van~Thourhout}, \citenamefont
  {Baets},\ and\ \citenamefont {Morthier}}]{Liu:2010}%
  \BibitemOpen
  \bibfield  {author} {\bibinfo {author} {\bibfnamefont {L.}~\bibnamefont
  {Liu}}, \bibinfo {author} {\bibfnamefont {R.}~\bibnamefont {Kumar}}, \bibinfo
  {author} {\bibfnamefont {K.}~\bibnamefont {Huybrechts}}, \bibinfo {author}
  {\bibfnamefont {T.}~\bibnamefont {Spuesens}}, \bibinfo {author}
  {\bibfnamefont {G.}~\bibnamefont {Roelkens}}, \bibinfo {author}
  {\bibfnamefont {E.-J.}\ \bibnamefont {Geluk}}, \bibinfo {author}
  {\bibfnamefont {T.}~\bibnamefont {de~Vries}}, \bibinfo {author}
  {\bibfnamefont {P.}~\bibnamefont {Regreny}}, \bibinfo {author} {\bibfnamefont
  {D.}~\bibnamefont {Van~Thourhout}}, \bibinfo {author} {\bibfnamefont
  {R.}~\bibnamefont {Baets}}, \ and\ \bibinfo {author} {\bibfnamefont
  {G.}~\bibnamefont {Morthier}},\ }\bibfield  {title} {\enquote {\bibinfo
  {title} {An ultra-small, low-power, all-optical flip-flop memory on a silicon
  chip},}\ }\href {http://dx.doi.org/10.1038/nphoton.2009.268} {\bibfield
  {journal} {\bibinfo  {journal} {Nat. Photonics}\ }\textbf {\bibinfo {volume}
  {4}},\ \bibinfo {pages} {182--187} (\bibinfo {year} {2010})}\BibitemShut
  {NoStop}%
\bibitem [{\citenamefont {Nozaki}\ \emph {et~al.}(2012)\citenamefont {Nozaki},
  \citenamefont {Shinya}, \citenamefont {Matsuo}, \citenamefont {Suzaki},
  \citenamefont {Segawa}, \citenamefont {Sato}, \citenamefont {Kawaguchi},
  \citenamefont {Takahashi},\ and\ \citenamefont {Notomi}}]{Nozaki:2012}%
  \BibitemOpen
  \bibfield  {author} {\bibinfo {author} {\bibfnamefont {K.}~\bibnamefont
  {Nozaki}}, \bibinfo {author} {\bibfnamefont {A.}~\bibnamefont {Shinya}},
  \bibinfo {author} {\bibfnamefont {S.}~\bibnamefont {Matsuo}}, \bibinfo
  {author} {\bibfnamefont {Y.}~\bibnamefont {Suzaki}}, \bibinfo {author}
  {\bibfnamefont {T.}~\bibnamefont {Segawa}}, \bibinfo {author} {\bibfnamefont
  {T.}~\bibnamefont {Sato}}, \bibinfo {author} {\bibfnamefont {Y.}~\bibnamefont
  {Kawaguchi}}, \bibinfo {author} {\bibfnamefont {R.}~\bibnamefont
  {Takahashi}}, \ and\ \bibinfo {author} {\bibfnamefont {M.}~\bibnamefont
  {Notomi}},\ }\bibfield  {title} {\enquote {\bibinfo {title} {Ultralow-power
  all-optical RAM based on nanocavities},}\ }\href
  {http://dx.doi.org/10.1038/nphoton.2012.2} {\bibfield  {journal} {\bibinfo
  {journal} {Nat. Photonics}\ }\textbf {\bibinfo {volume} {6}},\ \bibinfo
  {pages} {248--252} (\bibinfo {year} {2012})}\BibitemShut {NoStop}%
\bibitem [{\citenamefont {Kuramochi}\ \emph {et~al.}(2014)\citenamefont
  {Kuramochi}, \citenamefont {Nozaki}, \citenamefont {Shinya}, \citenamefont
  {Takeda}, \citenamefont {Sato}, \citenamefont {Matsuo}, \citenamefont
  {Taniyama}, \citenamefont {Sumikura},\ and\ \citenamefont
  {Notomi}}]{Kuramochi:2014}%
  \BibitemOpen
  \bibfield  {author} {\bibinfo {author} {\bibfnamefont {E.}~\bibnamefont
  {Kuramochi}}, \bibinfo {author} {\bibfnamefont {K.}~\bibnamefont {Nozaki}},
  \bibinfo {author} {\bibfnamefont {A.}~\bibnamefont {Shinya}}, \bibinfo
  {author} {\bibfnamefont {K.}~\bibnamefont {Takeda}}, \bibinfo {author}
  {\bibfnamefont {T.}~\bibnamefont {Sato}}, \bibinfo {author} {\bibfnamefont
  {S.}~\bibnamefont {Matsuo}}, \bibinfo {author} {\bibfnamefont
  {H.}~\bibnamefont {Taniyama}}, \bibinfo {author} {\bibfnamefont
  {H.}~\bibnamefont {Sumikura}}, \ and\ \bibinfo {author} {\bibfnamefont
  {M.}~\bibnamefont {Notomi}},\ }\bibfield  {title} {\enquote {\bibinfo {title}
  {Large-scale integration of wavelength-addressable all-optical memories on a
  photonic crystal chip},}\ }\href {http://dx.doi.org/10.1038/nphoton.2014.93}
  {\bibfield  {journal} {\bibinfo  {journal} {Nat. Photonics}\ }\textbf
  {\bibinfo {volume} {8}},\ \bibinfo {pages} {474--481} (\bibinfo {year}
  {2014})}\BibitemShut {NoStop}%
\bibitem [{\citenamefont {Gaufr\`{e}s}\ \emph {et~al.}(2012)\citenamefont
  {Gaufr\`{e}s}, \citenamefont {Izard}, \citenamefont {Noury}, \citenamefont
  {Le~Roux}, \citenamefont {Rasigade}, \citenamefont {Beck},\ and\
  \citenamefont {Vivien}}]{Gaufres:2012}%
  \BibitemOpen
  \bibfield  {author} {\bibinfo {author} {\bibfnamefont {E.}~\bibnamefont
  {Gaufr\`{e}s}}, \bibinfo {author} {\bibfnamefont {N.}~\bibnamefont {Izard}},
  \bibinfo {author} {\bibfnamefont {A.}~\bibnamefont {Noury}}, \bibinfo
  {author} {\bibfnamefont {X.}~\bibnamefont {Le~Roux}}, \bibinfo {author}
  {\bibfnamefont {G.}~\bibnamefont {Rasigade}}, \bibinfo {author}
  {\bibfnamefont {A.}~\bibnamefont {Beck}}, \ and\ \bibinfo {author}
  {\bibfnamefont {L.}~\bibnamefont {Vivien}},\ }\bibfield  {title} {\enquote
  {\bibinfo {title} {Light emission in silicon from carbon nanotubes},}\ }\href
  {\doibase 10.1021/nn204924n} {\bibfield  {journal} {\bibinfo  {journal} {ACS
  Nano}\ }\textbf {\bibinfo {volume} {6}},\ \bibinfo {pages} {3813--3819}
  (\bibinfo {year} {2012})}\BibitemShut {NoStop}%
\bibitem [{\citenamefont {Imamura}\ \emph {et~al.}(2013)\citenamefont
  {Imamura}, \citenamefont {Watahiki}, \citenamefont {Miura}, \citenamefont
  {Shimada},\ and\ \citenamefont {Kato}}]{Imamura:2013}%
  \BibitemOpen
  \bibfield  {author} {\bibinfo {author} {\bibfnamefont {S.}~\bibnamefont
  {Imamura}}, \bibinfo {author} {\bibfnamefont {R.}~\bibnamefont {Watahiki}},
  \bibinfo {author} {\bibfnamefont {R.}~\bibnamefont {Miura}}, \bibinfo
  {author} {\bibfnamefont {T.}~\bibnamefont {Shimada}}, \ and\ \bibinfo
  {author} {\bibfnamefont {Y.~K.}\ \bibnamefont {Kato}},\ }\bibfield  {title}
  {\enquote {\bibinfo {title} {Optical control of individual carbon nanotube
  light emitters by spectral double resonance in silicon microdisk
  resonators},}\ }\href {\doibase 10.1063/1.4802930} {\bibfield  {journal}
  {\bibinfo  {journal} {Appl. Phys. Lett.}\ }\textbf {\bibinfo {volume}
  {102}},\ \bibinfo {eid} {161102} (\bibinfo {year} {2013})}\BibitemShut
  {NoStop}%
\bibitem [{\citenamefont {Miura}\ \emph {et~al.}(2014)\citenamefont {Miura},
  \citenamefont {Imamura}, \citenamefont {Ohta}, \citenamefont {Ishii},
  \citenamefont {Liu}, \citenamefont {Shimada}, \citenamefont {Iwamoto},
  \citenamefont {Arakawa},\ and\ \citenamefont {Kato}}]{Miura:2014}%
  \BibitemOpen
  \bibfield  {author} {\bibinfo {author} {\bibfnamefont {R.}~\bibnamefont
  {Miura}}, \bibinfo {author} {\bibfnamefont {S.}~\bibnamefont {Imamura}},
  \bibinfo {author} {\bibfnamefont {R.}~\bibnamefont {Ohta}}, \bibinfo {author}
  {\bibfnamefont {A.}~\bibnamefont {Ishii}}, \bibinfo {author} {\bibfnamefont
  {X.}~\bibnamefont {Liu}}, \bibinfo {author} {\bibfnamefont {T.}~\bibnamefont
  {Shimada}}, \bibinfo {author} {\bibfnamefont {S.}~\bibnamefont {Iwamoto}},
  \bibinfo {author} {\bibfnamefont {Y.}~\bibnamefont {Arakawa}}, \ and\
  \bibinfo {author} {\bibfnamefont {Y.~K.}\ \bibnamefont {Kato}},\ }\bibfield
  {title} {\enquote {\bibinfo {title} {Ultralow mode-volume photonic crystal
  nanobeam cavities for high-efficiency coupling to individual carbon nanotube
  emitters},}\ }\href {http://dx.doi.org/10.1038/ncomms6580} {\bibfield
  {journal} {\bibinfo  {journal} {Nat. Commun.}\ }\textbf {\bibinfo {volume}
  {5}},\ \bibinfo {pages} {5580} (\bibinfo {year} {2014})}\BibitemShut
  {NoStop}%
\bibitem [{\citenamefont {Ando}(1997)}]{Ando:1997}%
  \BibitemOpen
  \bibfield  {author} {\bibinfo {author} {\bibfnamefont {T.}~\bibnamefont
  {Ando}},\ }\bibfield  {title} {\enquote {\bibinfo {title} {Excitons in carbon
  nanotubes},}\ }\href {\doibase 10.1143/JPSJ.66.1066} {\bibfield  {journal}
  {\bibinfo  {journal} {J. Phys. Soc. Jpn.}\ }\textbf {\bibinfo {volume}
  {66}},\ \bibinfo {pages} {1066--1073} (\bibinfo {year} {1997})}\BibitemShut
  {NoStop}%
\bibitem [{\citenamefont {Wang}\ \emph {et~al.}(2005)\citenamefont {Wang},
  \citenamefont {Dukovic}, \citenamefont {Brus},\ and\ \citenamefont
  {Heinz}}]{Wang:2005}%
  \BibitemOpen
  \bibfield  {author} {\bibinfo {author} {\bibfnamefont {F.}~\bibnamefont
  {Wang}}, \bibinfo {author} {\bibfnamefont {G.}~\bibnamefont {Dukovic}},
  \bibinfo {author} {\bibfnamefont {L.~E.}\ \bibnamefont {Brus}}, \ and\
  \bibinfo {author} {\bibfnamefont {T.~F.}\ \bibnamefont {Heinz}},\ }\bibfield
  {title} {\enquote {\bibinfo {title} {The optical resonances in carbon
  nanotubes arise from excitons},}\ }\href {\doibase 10.1126/science.1110265}
  {\bibfield  {journal} {\bibinfo  {journal} {Science}\ }\textbf {\bibinfo
  {volume} {308}},\ \bibinfo {pages} {838--841} (\bibinfo {year}
  {2005})}\BibitemShut {NoStop}%
\bibitem [{\citenamefont {Maultzsch}\ \emph {et~al.}(2005)\citenamefont
  {Maultzsch}, \citenamefont {Pomraenke}, \citenamefont {Reich}, \citenamefont
  {Chang}, \citenamefont {Prezzi}, \citenamefont {Ruini}, \citenamefont
  {Molinari}, \citenamefont {Strano}, \citenamefont {Thomsen},\ and\
  \citenamefont {Lienau}}]{Maultzsch:2005}%
  \BibitemOpen
  \bibfield  {author} {\bibinfo {author} {\bibfnamefont {J.}~\bibnamefont
  {Maultzsch}}, \bibinfo {author} {\bibfnamefont {R.}~\bibnamefont
  {Pomraenke}}, \bibinfo {author} {\bibfnamefont {S.}~\bibnamefont {Reich}},
  \bibinfo {author} {\bibfnamefont {E.}~\bibnamefont {Chang}}, \bibinfo
  {author} {\bibfnamefont {D.}~\bibnamefont {Prezzi}}, \bibinfo {author}
  {\bibfnamefont {A.}~\bibnamefont {Ruini}}, \bibinfo {author} {\bibfnamefont
  {E.}~\bibnamefont {Molinari}}, \bibinfo {author} {\bibfnamefont {M.~S.}\
  \bibnamefont {Strano}}, \bibinfo {author} {\bibfnamefont {C.}~\bibnamefont
  {Thomsen}}, \ and\ \bibinfo {author} {\bibfnamefont {C.}~\bibnamefont
  {Lienau}},\ }\bibfield  {title} {\enquote {\bibinfo {title} {Exciton binding
  energies in carbon nanotubes from two-photon photoluminescence},}\ }\href
  {\doibase 10.1103/PhysRevB.72.241402} {\bibfield  {journal} {\bibinfo
  {journal} {Phys. Rev. B}\ }\textbf {\bibinfo {volume} {72}},\ \bibinfo
  {pages} {241402} (\bibinfo {year} {2005})}\BibitemShut {NoStop}%
\bibitem [{\citenamefont {Weisman}\ and\ \citenamefont
  {Bachilo}(2003)}]{Weisman:2003}%
  \BibitemOpen
  \bibfield  {author} {\bibinfo {author} {\bibfnamefont {R.~B.}\ \bibnamefont
  {Weisman}}\ and\ \bibinfo {author} {\bibfnamefont {S.~M.}\ \bibnamefont
  {Bachilo}},\ }\bibfield  {title} {\enquote {\bibinfo {title} {Dependence of
  optical transition energies on structure for single-walled carbon nanotubes
  in aqueous suspension:遯ｶ・ｽ an empirical {K}ataura plot},}\ }\href {\doibase
  10.1021/nl034428i} {\bibfield  {journal} {\bibinfo  {journal} {Nano Lett.}\
  }\textbf {\bibinfo {volume} {3}},\ \bibinfo {pages} {1235--1238} (\bibinfo
  {year} {2003})}\BibitemShut {NoStop}%
\bibitem [{\citenamefont {Ajiki}\ and\ \citenamefont
  {Ando}(1994)}]{Ajiki:1994}%
  \BibitemOpen
  \bibfield  {author} {\bibinfo {author} {\bibfnamefont {H.}~\bibnamefont
  {Ajiki}}\ and\ \bibinfo {author} {\bibfnamefont {T.}~\bibnamefont {Ando}},\
  }\bibfield  {title} {\enquote {\bibinfo {title} {Aharonov-Bohm effect in
  carbon nanotubes},}\ }\href {\doibase 10.1016/0921-4526(94)91112-6}
  {\bibfield  {journal} {\bibinfo  {journal} {Physica B: Condensed Matter}\
  }\textbf {\bibinfo {volume} {201}},\ \bibinfo {pages} {349--352} (\bibinfo
  {year} {1994})}\BibitemShut {NoStop}%
\bibitem [{\citenamefont {Liu}\ \emph {et~al.}(2013)\citenamefont {Liu},
  \citenamefont {Hong}, \citenamefont {Zhou}, \citenamefont {Jin},
  \citenamefont {Li}, \citenamefont {Zhou}, \citenamefont {Liu}, \citenamefont
  {Wang}, \citenamefont {Zettl},\ and\ \citenamefont {Wang}}]{Liu:2013}%
  \BibitemOpen
  \bibfield  {author} {\bibinfo {author} {\bibfnamefont {K.}~\bibnamefont
  {Liu}}, \bibinfo {author} {\bibfnamefont {X.}~\bibnamefont {Hong}}, \bibinfo
  {author} {\bibfnamefont {Q.}~\bibnamefont {Zhou}}, \bibinfo {author}
  {\bibfnamefont {C.}~\bibnamefont {Jin}}, \bibinfo {author} {\bibfnamefont
  {J.}~\bibnamefont {Li}}, \bibinfo {author} {\bibfnamefont {W.}~\bibnamefont
  {Zhou}}, \bibinfo {author} {\bibfnamefont {J.}~\bibnamefont {Liu}}, \bibinfo
  {author} {\bibfnamefont {E.}~\bibnamefont {Wang}}, \bibinfo {author}
  {\bibfnamefont {A.}~\bibnamefont {Zettl}}, \ and\ \bibinfo {author}
  {\bibfnamefont {F.}~\bibnamefont {Wang}},\ }\bibfield  {title} {\enquote
  {\bibinfo {title} {High-throughput optical imaging and spectroscopy of
  individual carbon nanotubes in devices},}\ }\href
  {http://dx.doi.org/10.1038/nnano.2013.227} {\bibfield  {journal} {\bibinfo
  {journal} {Nat. Nanotech.}\ }\textbf {\bibinfo {volume} {8}},\ \bibinfo
  {pages} {917--922} (\bibinfo {year} {2013})}\BibitemShut {NoStop}%
\bibitem [{\citenamefont {Ohno}\ \emph {et~al.}(2006)\citenamefont {Ohno},
  \citenamefont {Iwasaki}, \citenamefont {Murakami}, \citenamefont {Kishimoto},
  \citenamefont {Maruyama},\ and\ \citenamefont {Mizutani}}]{Ohno:2006}%
  \BibitemOpen
  \bibfield  {author} {\bibinfo {author} {\bibfnamefont {Y.}~\bibnamefont
  {Ohno}}, \bibinfo {author} {\bibfnamefont {S.}~\bibnamefont {Iwasaki}},
  \bibinfo {author} {\bibfnamefont {Y.}~\bibnamefont {Murakami}}, \bibinfo
  {author} {\bibfnamefont {S.}~\bibnamefont {Kishimoto}}, \bibinfo {author}
  {\bibfnamefont {S.}~\bibnamefont {Maruyama}}, \ and\ \bibinfo {author}
  {\bibfnamefont {T.}~\bibnamefont {Mizutani}},\ }\bibfield  {title} {\enquote
  {\bibinfo {title} {Chirality-dependent environmental effects in
  photoluminescence of single-walled carbon nanotubes},}\ }\href {\doibase
  10.1103/PhysRevB.73.235427} {\bibfield  {journal} {\bibinfo  {journal} {Phys.
  Rev. B}\ }\textbf {\bibinfo {volume} {73}},\ \bibinfo {pages} {235427}
  (\bibinfo {year} {2006})}\BibitemShut {NoStop}%
\bibitem [{\citenamefont {Miyauchi}\ \emph {et~al.}(2007)\citenamefont
  {Miyauchi}, \citenamefont {Saito}, \citenamefont {Sato}, \citenamefont
  {Ohno}, \citenamefont {Iwasaki}, \citenamefont {Mizutani}, \citenamefont
  {Jiang},\ and\ \citenamefont {Maruyama}}]{Miyauchi:2007}%
  \BibitemOpen
  \bibfield  {author} {\bibinfo {author} {\bibfnamefont {Y.}~\bibnamefont
  {Miyauchi}}, \bibinfo {author} {\bibfnamefont {R.}~\bibnamefont {Saito}},
  \bibinfo {author} {\bibfnamefont {K.}~\bibnamefont {Sato}}, \bibinfo {author}
  {\bibfnamefont {Y.}~\bibnamefont {Ohno}}, \bibinfo {author} {\bibfnamefont
  {S.}~\bibnamefont {Iwasaki}}, \bibinfo {author} {\bibfnamefont
  {T.}~\bibnamefont {Mizutani}}, \bibinfo {author} {\bibfnamefont
  {J.}~\bibnamefont {Jiang}}, \ and\ \bibinfo {author} {\bibfnamefont
  {S.}~\bibnamefont {Maruyama}},\ }\bibfield  {title} {\enquote {\bibinfo
  {title} {Dependence of exciton transition energy of single-walled carbon
  nanotubes on surrounding dielectric materials},}\ }\href
  {http://www.sciencedirect.com/science/article/pii/S0009261407007464}
  {\bibfield  {journal} {\bibinfo  {journal} {Chem. Phys. Lett.}\ }\textbf
  {\bibinfo {volume} {442}},\ \bibinfo {pages} {394--399} (\bibinfo {year}
  {2007})}\BibitemShut {NoStop}%
\bibitem [{\citenamefont {Finnie}, \citenamefont {Homma},\ and\ \citenamefont
  {Lefebvre}(2005)}]{Finnie:2005}%
  \BibitemOpen
  \bibfield  {author} {\bibinfo {author} {\bibfnamefont {P.}~\bibnamefont
  {Finnie}}, \bibinfo {author} {\bibfnamefont {Y.}~\bibnamefont {Homma}}, \
  and\ \bibinfo {author} {\bibfnamefont {J.}~\bibnamefont {Lefebvre}},\
  }\bibfield  {title} {\enquote {\bibinfo {title} {Band-gap shift transition in
  the photoluminescence of single-walled carbon nanotubes},}\ }\href {\doibase
  10.1103/PhysRevLett.94.247401} {\bibfield  {journal} {\bibinfo  {journal}
  {Phys. Rev. Lett.}\ }\textbf {\bibinfo {volume} {94}},\ \bibinfo {pages}
  {247401} (\bibinfo {year} {2005})}\BibitemShut {NoStop}%
\bibitem [{\citenamefont {Lefebvre}\ and\ \citenamefont
  {Finnie}(2008)}]{Lefebvre:2008}%
  \BibitemOpen
  \bibfield  {author} {\bibinfo {author} {\bibfnamefont {J.}~\bibnamefont
  {Lefebvre}}\ and\ \bibinfo {author} {\bibfnamefont {P.}~\bibnamefont
  {Finnie}},\ }\bibfield  {title} {\enquote {\bibinfo {title} {Excited
  excitonic states in single-walled carbon nanotubes},}\ }\href {\doibase
  10.1021/nl080518h} {\bibfield  {journal} {\bibinfo  {journal} {Nano Lett.}\
  }\textbf {\bibinfo {volume} {8}},\ \bibinfo {pages} {1890--1895} (\bibinfo
  {year} {2008})}\BibitemShut {NoStop}%
\bibitem [{\citenamefont {Homma}\ \emph {et~al.}(2013)\citenamefont {Homma},
  \citenamefont {Chiashi}, \citenamefont {Yamamoto}, \citenamefont {Kono},
  \citenamefont {Matsumoto}, \citenamefont {Shitaba},\ and\ \citenamefont
  {Sato}}]{Homma:2013}%
  \BibitemOpen
  \bibfield  {author} {\bibinfo {author} {\bibfnamefont {Y.}~\bibnamefont
  {Homma}}, \bibinfo {author} {\bibfnamefont {S.}~\bibnamefont {Chiashi}},
  \bibinfo {author} {\bibfnamefont {T.}~\bibnamefont {Yamamoto}}, \bibinfo
  {author} {\bibfnamefont {K.}~\bibnamefont {Kono}}, \bibinfo {author}
  {\bibfnamefont {D.}~\bibnamefont {Matsumoto}}, \bibinfo {author}
  {\bibfnamefont {J.}~\bibnamefont {Shitaba}}, \ and\ \bibinfo {author}
  {\bibfnamefont {S.}~\bibnamefont {Sato}},\ }\bibfield  {title} {\enquote
  {\bibinfo {title} {Photoluminescence measurements and molecular dynamics
  simulations of water adsorption on the hydrophobic surface of a carbon
  nanotube in water vapor},}\ }\href {\doibase 10.1103/PhysRevLett.110.157402}
  {\bibfield  {journal} {\bibinfo  {journal} {Phys. Rev. Lett.}\ }\textbf
  {\bibinfo {volume} {110}},\ \bibinfo {pages} {157402} (\bibinfo {year}
  {2013})}\BibitemShut {NoStop}%
\bibitem [{\citenamefont {Xiao}, \citenamefont {Anderson},\ and\ \citenamefont
  {Fraser}(2014)}]{Xiao:2014}%
  \BibitemOpen
  \bibfield  {author} {\bibinfo {author} {\bibfnamefont {Y.-F.}\ \bibnamefont
  {Xiao}}, \bibinfo {author} {\bibfnamefont {M.~D.}\ \bibnamefont {Anderson}},
  \ and\ \bibinfo {author} {\bibfnamefont {J.~M.}\ \bibnamefont {Fraser}},\
  }\bibfield  {title} {\enquote {\bibinfo {title} {Photoluminescence saturation
  independent of excitation pathway in air-suspended single-walled carbon
  nanotubes},}\ }\href {\doibase 10.1103/PhysRevB.89.235440} {\bibfield
  {journal} {\bibinfo  {journal} {Phys. Rev. B}\ }\textbf {\bibinfo {volume}
  {89}},\ \bibinfo {pages} {235440} (\bibinfo {year} {2014})}\BibitemShut
  {NoStop}%
\bibitem [{\citenamefont {Milkie}\ \emph {et~al.}(2005)\citenamefont {Milkie},
  \citenamefont {Staii}, \citenamefont {Paulson}, \citenamefont {Hindman},
  \citenamefont {Johnson},\ and\ \citenamefont {Kikkawa}}]{Milkie:2005}%
  \BibitemOpen
  \bibfield  {author} {\bibinfo {author} {\bibfnamefont {D.~E.}\ \bibnamefont
  {Milkie}}, \bibinfo {author} {\bibfnamefont {C.}~\bibnamefont {Staii}},
  \bibinfo {author} {\bibfnamefont {S.}~\bibnamefont {Paulson}}, \bibinfo
  {author} {\bibfnamefont {E.}~\bibnamefont {Hindman}}, \bibinfo {author}
  {\bibfnamefont {A.~T.}\ \bibnamefont {Johnson}}, \ and\ \bibinfo {author}
  {\bibfnamefont {J.~M.}\ \bibnamefont {Kikkawa}},\ }\bibfield  {title}
  {\enquote {\bibinfo {title} {Controlled switching of optical emission
  energies in semiconducting single-walled carbon nanotubes},}\ }\href
  {\doibase 10.1021/nl050688j} {\bibfield  {journal} {\bibinfo  {journal} {Nano
  Lett.}\ }\textbf {\bibinfo {volume} {5}},\ \bibinfo {pages} {1135--1138}
  (\bibinfo {year} {2005})}\BibitemShut {NoStop}%
\bibitem [{\citenamefont {Moritsubo}\ \emph {et~al.}(2010)\citenamefont
  {Moritsubo}, \citenamefont {Murai}, \citenamefont {Shimada}, \citenamefont
  {Murakami}, \citenamefont {Chiashi}, \citenamefont {Maruyama},\ and\
  \citenamefont {Kato}}]{Moritsubo:2010}%
  \BibitemOpen
  \bibfield  {author} {\bibinfo {author} {\bibfnamefont {S.}~\bibnamefont
  {Moritsubo}}, \bibinfo {author} {\bibfnamefont {T.}~\bibnamefont {Murai}},
  \bibinfo {author} {\bibfnamefont {T.}~\bibnamefont {Shimada}}, \bibinfo
  {author} {\bibfnamefont {Y.}~\bibnamefont {Murakami}}, \bibinfo {author}
  {\bibfnamefont {S.}~\bibnamefont {Chiashi}}, \bibinfo {author} {\bibfnamefont
  {S.}~\bibnamefont {Maruyama}}, \ and\ \bibinfo {author} {\bibfnamefont
  {Y.~K.}\ \bibnamefont {Kato}},\ }\bibfield  {title} {\enquote {\bibinfo
  {title} {Exciton diffusion in air-suspended single-walled carbon
  nanotubes},}\ }\href {\doibase 10.1103/PhysRevLett.104.247402} {\bibfield
  {journal} {\bibinfo  {journal} {Phys. Rev. Lett.}\ }\textbf {\bibinfo
  {volume} {104}},\ \bibinfo {pages} {247402} (\bibinfo {year}
  {2010})}\BibitemShut {NoStop}%
\bibitem [{\citenamefont {Maruyama}\ \emph {et~al.}(2002)\citenamefont
  {Maruyama}, \citenamefont {Kojima}, \citenamefont {Miyauchi}, \citenamefont
  {Chiashi},\ and\ \citenamefont {Kohno}}]{Maruyama:2002}%
  \BibitemOpen
  \bibfield  {author} {\bibinfo {author} {\bibfnamefont {S.}~\bibnamefont
  {Maruyama}}, \bibinfo {author} {\bibfnamefont {R.}~\bibnamefont {Kojima}},
  \bibinfo {author} {\bibfnamefont {Y.}~\bibnamefont {Miyauchi}}, \bibinfo
  {author} {\bibfnamefont {S.}~\bibnamefont {Chiashi}}, \ and\ \bibinfo
  {author} {\bibfnamefont {M.}~\bibnamefont {Kohno}},\ }\bibfield  {title}
  {\enquote {\bibinfo {title} {Low-temperature synthesis of high-purity
  single-walled carbon nanotubes from alcohol},}\ }\href
  {http://www.sciencedirect.com/science/article/pii/S0009261402008382}
  {\bibfield  {journal} {\bibinfo  {journal} {Chem. Phys. Lett.}\ }\textbf
  {\bibinfo {volume} {360}},\ \bibinfo {pages} {229--234} (\bibinfo {year}
  {2002})}\BibitemShut {NoStop}%
\bibitem [{\citenamefont {Jiang}\ \emph {et~al.}(2015)\citenamefont {Jiang},
  \citenamefont {Kumamoto}, \citenamefont {Ishii}, \citenamefont {Yoshida},
  \citenamefont {Shimada},\ and\ \citenamefont {Kato}}]{Jiang:2015}%
  \BibitemOpen
  \bibfield  {author} {\bibinfo {author} {\bibfnamefont {M.}~\bibnamefont
  {Jiang}}, \bibinfo {author} {\bibfnamefont {Y.}~\bibnamefont {Kumamoto}},
  \bibinfo {author} {\bibfnamefont {A.}~\bibnamefont {Ishii}}, \bibinfo
  {author} {\bibfnamefont {M.}~\bibnamefont {Yoshida}}, \bibinfo {author}
  {\bibfnamefont {T.}~\bibnamefont {Shimada}}, \ and\ \bibinfo {author}
  {\bibfnamefont {Y.~K.}\ \bibnamefont {Kato}},\ }\bibfield  {title} {\enquote
  {\bibinfo {title} {Gate-controlled generation of optical pulse trains using
  individual carbon nanotubes},}\ }\href {http://dx.doi.org/10.1038/ncomms7335}
  {\bibfield  {journal} {\bibinfo  {journal} {Nat. Commun.}\ }\textbf {\bibinfo
  {volume} {6}},\ \bibinfo {pages} {6335} (\bibinfo {year} {2015})}\BibitemShut
  {NoStop}%
\bibitem [{\citenamefont {Ishii}, \citenamefont {Yoshida},\ and\ \citenamefont
  {Kato}(2015)}]{Ishii:2015}%
  \BibitemOpen
  \bibfield  {author} {\bibinfo {author} {\bibfnamefont {A.}~\bibnamefont
  {Ishii}}, \bibinfo {author} {\bibfnamefont {M.}~\bibnamefont {Yoshida}}, \
  and\ \bibinfo {author} {\bibfnamefont {Y.~K.}\ \bibnamefont {Kato}},\
  }\bibfield  {title} {\enquote {\bibinfo {title} {Exciton diffusion, end
  quenching, and exciton-exciton annihilation in individual air-suspended
  carbon nanotubes},}\ }\href {\doibase 10.1103/PhysRevB.91.125427} {\bibfield
  {journal} {\bibinfo  {journal} {Phys. Rev. B}\ }\textbf {\bibinfo {volume}
  {91}},\ \bibinfo {pages} {125427} (\bibinfo {year} {2015})}\BibitemShut
  {NoStop}%
\bibitem [{\citenamefont {Georgi}\ \emph {et~al.}(2008)\citenamefont {Georgi},
  \citenamefont {Hartmann}, \citenamefont {Gokus}, \citenamefont {Green},
  \citenamefont {Hersam},\ and\ \citenamefont {Hartschuh}}]{Georgi:2008}%
  \BibitemOpen
  \bibfield  {author} {\bibinfo {author} {\bibfnamefont {C.}~\bibnamefont
  {Georgi}}, \bibinfo {author} {\bibfnamefont {N.}~\bibnamefont {Hartmann}},
  \bibinfo {author} {\bibfnamefont {T.}~\bibnamefont {Gokus}}, \bibinfo
  {author} {\bibfnamefont {A.~A.}\ \bibnamefont {Green}}, \bibinfo {author}
  {\bibfnamefont {M.~C.}\ \bibnamefont {Hersam}}, \ and\ \bibinfo {author}
  {\bibfnamefont {A.}~\bibnamefont {Hartschuh}},\ }\bibfield  {title} {\enquote
  {\bibinfo {title} {Photoinduced luminescence blinking and bleaching in
  individual single-walled carbon nanotubes},}\ }\href {\doibase
  10.1002/cphc.200800179} {\bibfield  {journal} {\bibinfo  {journal} {Chem.
  Phys. Chem.}\ }\textbf {\bibinfo {volume} {9}},\ \bibinfo {pages}
  {1460--1464} (\bibinfo {year} {2008})}\BibitemShut {NoStop}%
\bibitem [{\citenamefont {Yoshikawa}, \citenamefont {Matsuda},\ and\
  \citenamefont {Kanemitsu}(2010)}]{Yoshikawa:2010}%
  \BibitemOpen
  \bibfield  {author} {\bibinfo {author} {\bibfnamefont {K.}~\bibnamefont
  {Yoshikawa}}, \bibinfo {author} {\bibfnamefont {K.}~\bibnamefont {Matsuda}},
  \ and\ \bibinfo {author} {\bibfnamefont {Y.}~\bibnamefont {Kanemitsu}},\
  }\bibfield  {title} {\enquote {\bibinfo {title} {Exciton transport in
  suspended single carbon nanotubes studied by photoluminescence imaging
  spectroscopy},}\ }\href {\doibase 10.1021/jp911518h} {\bibfield  {journal}
  {\bibinfo  {journal} {J. Phys. Chem. C}\ }\textbf {\bibinfo {volume} {114}},\
  \bibinfo {pages} {4353--4356} (\bibinfo {year} {2010})}\BibitemShut {NoStop}%
\bibitem [{\citenamefont {Lefebvre}, \citenamefont {Finnie},\ and\
  \citenamefont {Homma}(2004)}]{Lefebvre:2004}%
  \BibitemOpen
  \bibfield  {author} {\bibinfo {author} {\bibfnamefont {J.}~\bibnamefont
  {Lefebvre}}, \bibinfo {author} {\bibfnamefont {P.}~\bibnamefont {Finnie}}, \
  and\ \bibinfo {author} {\bibfnamefont {Y.}~\bibnamefont {Homma}},\ }\bibfield
   {title} {\enquote {\bibinfo {title} {Temperature-dependent photoluminescence
  from single-walled carbon nanotubes},}\ }\href {\doibase
  10.1103/PhysRevB.70.045419} {\bibfield  {journal} {\bibinfo  {journal} {Phys.
  Rev. B}\ }\textbf {\bibinfo {volume} {70}},\ \bibinfo {pages} {045419}
  (\bibinfo {year} {2004})}\BibitemShut {NoStop}%
\bibitem [{\citenamefont {Matsuda}\ \emph {et~al.}(2008)\citenamefont
  {Matsuda}, \citenamefont {Inoue}, \citenamefont {Murakami}, \citenamefont
  {Maruyama},\ and\ \citenamefont {Kanemitsu}}]{Matsuda:2008_2}%
  \BibitemOpen
  \bibfield  {author} {\bibinfo {author} {\bibfnamefont {K.}~\bibnamefont
  {Matsuda}}, \bibinfo {author} {\bibfnamefont {T.}~\bibnamefont {Inoue}},
  \bibinfo {author} {\bibfnamefont {Y.}~\bibnamefont {Murakami}}, \bibinfo
  {author} {\bibfnamefont {S.}~\bibnamefont {Maruyama}}, \ and\ \bibinfo
  {author} {\bibfnamefont {Y.}~\bibnamefont {Kanemitsu}},\ }\bibfield  {title}
  {\enquote {\bibinfo {title} {Exciton dephasing and multiexciton
  recombinations in a single carbon nanotube},}\ }\href {\doibase
  10.1103/PhysRevB.77.033406} {\bibfield  {journal} {\bibinfo  {journal} {Phys.
  Rev. B}\ }\textbf {\bibinfo {volume} {77}},\ \bibinfo {pages} {033406}
  (\bibinfo {year} {2008})}\BibitemShut {NoStop}%
\bibitem [{\citenamefont {Yoshikawa}\ \emph {et~al.}(2009)\citenamefont
  {Yoshikawa}, \citenamefont {Matsunaga}, \citenamefont {Matsuda},\ and\
  \citenamefont {Kanemitsu}}]{Yoshikawa:2009}%
  \BibitemOpen
  \bibfield  {author} {\bibinfo {author} {\bibfnamefont {K.}~\bibnamefont
  {Yoshikawa}}, \bibinfo {author} {\bibfnamefont {R.}~\bibnamefont
  {Matsunaga}}, \bibinfo {author} {\bibfnamefont {K.}~\bibnamefont {Matsuda}},
  \ and\ \bibinfo {author} {\bibfnamefont {Y.}~\bibnamefont {Kanemitsu}},\
  }\bibfield  {title} {\enquote {\bibinfo {title} {Mechanism of exciton
  dephasing in a single carbon nanotube studied by photoluminescence
  spectroscopy},}\ }\href {\doibase 10.1063/1.3089843} {\bibfield  {journal}
  {\bibinfo  {journal} {Appl. Phys. Lett.}\ }\textbf {\bibinfo {volume} {94}},\
  \bibinfo {pages} {093109} (\bibinfo {year} {2009})}\BibitemShut {NoStop}%
\bibitem [{\citenamefont {Adamson}(1990)}]{Adamson:1990}%
  \BibitemOpen
  \bibfield  {author} {\bibinfo {author} {\bibfnamefont {A.~W.}\ \bibnamefont
  {Adamson}},\ }\href@noop {} {\emph {\bibinfo {title} {Physical chemistry of
  surfaces}}}\ (\bibinfo  {publisher} {Wiley},\ \bibinfo {address} {New York},\
  \bibinfo {year} {1990})\BibitemShut {NoStop}%
\bibitem [{\citenamefont {Berber}, \citenamefont {Kwon},\ and\ \citenamefont
  {Tom\'{a}nek}(2000)}]{Berber:2000}%
  \BibitemOpen
  \bibfield  {author} {\bibinfo {author} {\bibfnamefont {S.}~\bibnamefont
  {Berber}}, \bibinfo {author} {\bibfnamefont {Y.-K.}\ \bibnamefont {Kwon}}, \
  and\ \bibinfo {author} {\bibfnamefont {D.}~\bibnamefont {Tom\'{a}nek}},\
  }\bibfield  {title} {\enquote {\bibinfo {title} {Unusually high thermal
  conductivity of carbon nanotubes},}\ }\href
  {https://link.aps.org/doi/10.1103/PhysRevLett.84.4613} {\bibfield  {journal}
  {\bibinfo  {journal} {Phys. Rev. Lett.}\ }\textbf {\bibinfo {volume} {84}},\
  \bibinfo {pages} {4613--4616} (\bibinfo {year} {2000})}\BibitemShut {NoStop}%
\bibitem [{\citenamefont {Fujii}\ \emph {et~al.}(2005)\citenamefont {Fujii},
  \citenamefont {Zhang}, \citenamefont {Xie}, \citenamefont {Ago},
  \citenamefont {Takahashi}, \citenamefont {Ikuta}, \citenamefont {Abe},\ and\
  \citenamefont {Shimizu}}]{Fujii:2005}%
  \BibitemOpen
  \bibfield  {author} {\bibinfo {author} {\bibfnamefont {M.}~\bibnamefont
  {Fujii}}, \bibinfo {author} {\bibfnamefont {X.}~\bibnamefont {Zhang}},
  \bibinfo {author} {\bibfnamefont {H.}~\bibnamefont {Xie}}, \bibinfo {author}
  {\bibfnamefont {H.}~\bibnamefont {Ago}}, \bibinfo {author} {\bibfnamefont
  {K.}~\bibnamefont {Takahashi}}, \bibinfo {author} {\bibfnamefont
  {T.}~\bibnamefont {Ikuta}}, \bibinfo {author} {\bibfnamefont
  {H.}~\bibnamefont {Abe}}, \ and\ \bibinfo {author} {\bibfnamefont
  {T.}~\bibnamefont {Shimizu}},\ }\bibfield  {title} {\enquote {\bibinfo
  {title} {Measuring the thermal conductivity of a single carbon nanotube},}\
  }\href {https://link.aps.org/doi/10.1103/PhysRevLett.95.065502} {\bibfield
  {journal} {\bibinfo  {journal} {Phys. Rev. Lett.}\ }\textbf {\bibinfo
  {volume} {95}},\ \bibinfo {pages} {065502} (\bibinfo {year}
  {2005})}\BibitemShut {NoStop}%
\bibitem [{\citenamefont {Yu}\ \emph {et~al.}(2005)\citenamefont {Yu},
  \citenamefont {Shi}, \citenamefont {Yao}, \citenamefont {Li},\ and\
  \citenamefont {Majumdar}}]{Yu:2005}%
  \BibitemOpen
  \bibfield  {author} {\bibinfo {author} {\bibfnamefont {C.}~\bibnamefont
  {Yu}}, \bibinfo {author} {\bibfnamefont {L.}~\bibnamefont {Shi}}, \bibinfo
  {author} {\bibfnamefont {Z.}~\bibnamefont {Yao}}, \bibinfo {author}
  {\bibfnamefont {D.}~\bibnamefont {Li}}, \ and\ \bibinfo {author}
  {\bibfnamefont {A.}~\bibnamefont {Majumdar}},\ }\bibfield  {title} {\enquote
  {\bibinfo {title} {Thermal conductance and thermopower of an individual
  single-wall carbon nanotube},}\ }\href {\doibase 10.1021/nl051044e}
  {\bibfield  {journal} {\bibinfo  {journal} {Nano Lett.}\ }\textbf {\bibinfo
  {volume} {5}},\ \bibinfo {pages} {1842--1846} (\bibinfo {year}
  {2005})}\BibitemShut {NoStop}%
\end{thebibliography}
\end{document}